\documentstyle [aps,twocolumn]{revtex}
\catcode`\@=11
\newcommand{\be}{\begin{equation}}
\newcommand{\ee}{\end{equation}}
\newcommand{\bea}{\begin{eqnarray}}
\newcommand{\eea}{\end{eqnarray}}

\newcommand{\p}{\partial}

\newcommand{\ri}{\mbox{i}}
\newcommand{\re}{\mbox{e}}

\begin {document}
%\begin{titlepage}
%\begin{flushright}
%\today
%\end{flushright}
%\vspace{0.5cm}
%\begin{center}
\title{New Chiral Universality Class in a Frustrated
Three-Leg Spin Ladder}
%\vspace{1.8cm}
%\vspace{0.5cm}
\vspace{2cm}

\author{P. Azaria$^1$,   P. Lecheminant$^2$,
  and A. A. Nersesyan$^{3,4}$}

\vspace{0.5cm}

\address{$^1$ Laboratoire de Physique Th\'eorique des Liquides,
Universit\'e Pierre et Marie Curie, 4 Place Jussieu, 75252 Paris,
France\\
$^2$ Laboratoire de Physique Th\'eorique et Mod\'elisation,
Universit\'e de Cergy-Pontoise, Site de Saint Martin,
2 avenue Adolphe Chauvin, 95302
Cergy-Pontoise Cedex, France\\
$^3$ Physikalisches Institut, Universit\"{a}t Bonn,
Nu\ss allee 12, 53115 Bonn, Germany\\
$^4$ Institute of Physics, Tamarashvili 6,
380077, Tbilisi, Georgia}

\vspace{3cm}

\address{\rm (Received: )}
\address{\mbox{ }}
\address{\parbox{14cm}{\rm \mbox{ }\mbox{ }
%\begin{abstract}  
%\par  
We study  a model of three
$S=1/2$ antiferromagnetic Heisenberg spin chains weakly coupled by
on-rung and plaquette-diagonal interchain
interactions. It is shown that
the model exhibits a critical phase
with central charge $C=2$ and belongs to the
class of ``chirally stabilized'' liquids
recently introduced by Andrei, Douglas, and Jerez.
By allowing anisotropic interactions in spin space,  we
find an exact solution at
a Toulouse point which captures all universal properties of the model,
including the SU(2) symmetric
case. At the new critical point the massless degrees of freedom are
described in terms of an effective $S = 1/2$ Heisenberg spin chain
and two critical Ising models.
We discuss the spectral properties of the model,
compute spin-spin correlation functions and estimate the NMR relaxation
rate.
}}
\address{\mbox{ }}
\address{\parbox{14cm}{\rm PACS No: 75.10.Jm, 75.40.Gb}}
\maketitle

\makeatletter
\global\@specialpagefalse
\makeatother
%\end{abstract}
%cond-mat/96
%\end{center}
%PACS numbers:
%\end{titlepage}
In close parallel with  qualitative difference between
the integer-spin and half-integer-spin antiferromagnetic chains,
first predicted by Haldane\cite{haldane}, the universal properties of
standard spin ladders, $i.e.$ those with the interchain
exchange interaction ($J_{\perp}$) across the rungs,
dramatically depend on the parity of the number of legs\cite{dagotto}.
Ladders with even number of legs are  disordered
spin liquids with a finite gap in the
excitation spectrum,
while
in odd-legged ladders
there exists a gapless branch in the spectrum
implying that the spin correlations
decay algebraically.
In the low-energy limit,
the critical behavior of the odd-legged ladders is characterized by the
central charge
$C = 1$ corresponding to
an effective $S = 1/2$  Heisenberg spin chain\cite{khve}.
The interesting question is
whether frustration can lead to new types of
infrared (IR) behavior not encountered
by the above described two ``even-odd'' scenarios.
The existence of real
quasi-1D antiferromagnets with a zigzag interchain interaction, such as
$Cs_2 Cu Cl_4$\cite{coldea} and $Cu_2 (C_5 H_{12} N_2) Cl_4$\cite{chab},
indicates that this question is not purely academic.
The role of frustration has recently been addressed
in a two-chain model with a weak zigzag interchain
coupling.
For an isotropic antiferromagnetic interaction the spectrum has
an exponentially small gap,  and the ground state is
spontaneously dimerized\cite{white,allen}.
 However,
an easy plane
XXZ-type anisotropy supports a gapless
phase characterized by nonzero local
spin currents polarized along the anisotropy axis and
algebraically decaying incommensurate spin correlations\cite{shura}.

Recently another type of frustrated two-leg ladders was
discussed\cite{oitmaa}.
In addition to the standard on-rung coupling $J_{\perp}$,
the new model
also includes an interaction $J_{\times}$ along both diagonals of
elementary plaquettes.
In this letter, we consider a three-chain generalization of such a model:
\bea
{\cal H} = \sum_n \{ J_{\parallel} \sum_{j=0,1,2} {\bf S}_{j,n}
\cdot {\bf S}_{j,n+1}
+
J_{\perp}  {\bf S}_{0,n}\cdot \left({\bf S}_{1,n}
+ {\bf S}_{2,n}\right) \nonumber \\
+ J_{\times}
\left[\left({\bf S}_{1,n} + {\bf S}_{2,n}\right)
\cdot {\bf S}_{0,n+1} +
\left({\bf S}_{1,n+1} + {\bf S}_{2,n+1}\right)
\cdot {\bf S}_{0,n}\right] \}
\label{hamiltres}
\eea
where frustration introduced by the interaction $J_{\times}$ shows up in
a highly nontrivial way.
We shall show that in a certain range of parameters the model displays
a new type of critical behavior in the IR limit,
with the  central charge $C=2$, rather  than $C = 1$ as in usual
(nonfrustrated)
three-leg ladders\cite{arri}. This critical behavior
is identified  as the universality class
of ``chirally stabilized'' fluids,
recently introduced by Andrei {\it et~ al.}\cite{andrei}.

We shall assume that
$0 < J_{\perp},J_{\times} \ll J_{\parallel}$. Under this condition
a continuum description can be adopted
in which the spin densities in each chain  are represented as
${\bf S}_j\left(x\right) = {\bf J}_j\left(x\right) +
 \left(-1\right)^{x/a_0}{\bf n}_j\left(x\right)$,
where ${\bf J}_j  = {\bf J}_{jR}  + {\bf J}_{jL}$
and ${\bf n}_j$ are, respectively, the smooth and staggered parts of
the magnetization.
In the continuum limit the Hamiltonian density of the original model
(\ref{hamiltres}) takes the form:
\bea
{\cal H} &=& \sum_{j=0,1,2} \frac{2\pi v_j}{3} \; \left({\bf J}_{jR} \cdot
{\bf J}_{jR} +
{\bf J}_{jL} \cdot {\bf J}_{jL} \right) \nonumber\\
&+&
g_1\; {\bf n}_0 \cdot \left({\bf n}_1 + {\bf n}_2\right)
+
g_2 \; {\bf J}_0 \cdot \left( {\bf J}_1 + {\bf J}_2\right).
\label{h-cont.lim}
\eea
Here the first line describes three decoupled chains in terms of three
critical SU(2)$_1$ Wess-Zumino-Novikov-Witten (WZNW) models\cite{affleck},
$v_j \sim J_{\parallel} a_0$ being the spin velocities. The second line in
(\ref{h-cont.lim}) includes the interchain coupling terms with the
constants:
$
g_1 = (J_{\perp}- 2 J_{\times})a_0, ~~g_2 = (J_{\perp}+ 2 J_{\times})a_0.
$
We stress that in the continuum limit there is
no marginally relevant twist perturbation,
${\bf n}_i \cdot \partial_x {\bf n}_j$,
which appears in the description of spin ladders
with small zigzag interchain coupling\cite{shura,azaria}.

The two interaction terms in (\ref{h-cont.lim}) are of different nature. 
The first one is a relevant perturbation with  scaling dimension
$d=1$. The second term describes
an interaction between the total current of the
surface chains,   
$
{\bf I} = {\bf J}_1 + {\bf J}_2
$
and that of the middle chain.
This interaction is only marginal and, as long as
$g_1$ is not too small, can be discarded.
As a result, for generic values of
$g_1$ and $g_2$, the low energy physics of our
model will be essentially that of the
standard three-leg ladder, and frustration will
plays no role
(except for renormalization of the mass gaps and
velocities\cite{shelton}).
The important point,
though, is that in contrast with non-frustrated ladders,  the two coupling
constants $g_1,g_2$ can vary independently, and
there exists a vicinity of the line $J_{\perp}= 2 J_{\times}$ ($g_1 = 0$)
where the low-energy properties of the model are mainly determined
by the
current-current interchain interaction.
Remarkably enough, exactly at $g_1 = 0$ an anisotropic version of this
model
is exactly solvable
at a Toulouse point. The solution describes a {\it new} fixed point
with a larger central
charge $C=2$ which, as we shall demonstrate,
govern the physics for sufficiently small $g_1$.
Our strategy will be first to
present our solution and discuss the physical properties of the system
at the special point $J_{\perp}= 2 J_{\times}$,
and then to explore
its neighborhood for  $g_1$ small.

{\bf Toulouse point solution}.
 We start by
neglecting the interaction between currents of the same chirality.  
Such terms lead
to renormalization of the velocities
of the excitations which can be effectively taken
into account by allowing
the surface chain velocities ($v_1 = v_2$) to be different from
that of the middle chain $v_0$.
In the noninteracting case ($g_2 = 0$),
the central charge of two surface chains is $C = 2$. On the other hand,
the total current ${\bf I}$, being the sum of two SU(2)$_1$ currents,
satisfies the SU(2)$_2$ Kac-Moody algebra.
Since the central charge of the SU(2)$_2$-symmetric WZNW model is $C =
3/2$,
some degrees of freedom
should account for the missing $C = 1/2$.
Those are associated with
a discrete (Z$_2$)  nonmagnetic
$1 \leftrightarrow 2$ interchange symmetry, and remain decoupled
and critical with the central charge $C = 1/2$.
The simplest way to see this  is to exploit the equivalence
SU(2) $\times$ SU(2) $\approx$ SO(4) and use the representation
of two  SU(2)$_1$ currents in terms of a
quadruplet of real (Majorana) fermions,
$\xi^{0}$ and
$\xi^{a}~ (a=1,2,3)$\cite{allen}: $ I_{\alpha}^a
= -\frac{\ri}{2} \; \epsilon^{abc}
\xi_{\alpha}^b \xi_{\alpha}^c, ~
\left(J_{\alpha 1} - J_{\alpha 2}\right)^a = \ri
\; \xi_{\alpha}^{a} \xi_{\alpha}^{0}$,
where $\alpha=R,L$.
Then one easily finds that the contribution of the massless Majorana
fermion $\xi^0$
decouples from the rest of the spectrum, and the
effective Hamiltonian at the special point $g_1 = 0$ reads:
\be
{\cal H} =
- \ri \frac{v_1}{2} \left(\xi_{R}^{0} \partial_x  \xi_{R}^{0} -
\xi_{L}^{0} \partial_x  \xi_{L}^{0}  \right)
+ \bar{\cal H} [{\bf I}, {\bf J}_0].
\label{hkondo}
\ee
All nontrivial physics is incorporated in the current dependent part of
the Hamiltonian, $\bar{\cal H}$, describing marginally coupled SU(2)$_2$
and SU(2)$_1$ WZNW models.
Notice that  $\bar{\cal H}$ in turn
separates into two commuting and
{\it chirally asymmetric} parts:
$\bar{\cal H} = {\cal H}_1 + {\cal H}_2, ~([{\cal H}_1, {\cal H}_2] = 0)$,
where
\be
{\cal H}_1 =  \frac{\pi v_1}{2} \;  {\bf I}_{R} \cdot{\bf I}_{R} +
 \frac{2\pi v_0}{3} \; {\bf J}_{0L} \cdot{\bf J}_{0L}
+ g_2 \;  {\bf I} _{R} \cdot {\bf J}_{0L},
\label{ham1}
\ee
with ${\cal H}_2$  obtained from ${\cal H}_1$ by
inverting chiralities of all the currents.
The Hamiltonian (\ref{ham1}) resembles the two-channel Kondo
model\cite{emery}:
in the latter case the SU(2)$_2$ current ${\bf I}_{R}$
describes  two-flavor spin excitations
of the conduction electrons, while the SU(2)$_1$ current ${\bf J}_{0L}$
is replaced by the local spin density of the impurity spin $S = 1/2$.
Moreover, one can show that (\ref{hkondo}) corresponds to  
the Hamiltonian of the two
channel Kondo lattice away from half-filling with a nearest
neighbour interaction between the impurity spins.

A simple renormalization group analysis reveals that, at $g_2 > 0$,
the interaction is marginally
relevant. Usually the development of
a strong coupling regime is accompanied by a
dynamical mass generation and the loss of conformal
invariance at the strong coupling
fixed point. This is
indeed the case for the model of two marginally coupled $S=1/2$ spin
chain with zigzag interaction\cite{white,allen}.
We shall see, however, that due to chiral asymmetry of $\bar{\cal H}$  
the effective interaction, as in the two-channel Kondo model,
flows towards an intermediate fixed point 
where conformal invariance is recovered with
a smaller central charge.

The model (\ref{ham1}) is Bethe-Ansatz solvable\cite{polyakov}
(see also \cite{andrei}).
Here we present an exact solution for a U(1) version of
the model, characterized by anisotropic interaction ($g_2 \rightarrow
g_{\parallel}, g_{\perp}$), which allows us to
investigate the spectrum of the model and
estimate  asymptotics of the correlation functions.
Our solution is based on a mapping onto Majorana fermions.
Using Abelian bosonization, we shall exploit the existence of a
Toulouse-like point where the fermions are free\cite{emery}.
We start with Abelian bosonization
of the SU(2)$_1$ current ${\bf J}_0$.
Introducing a massless bosonic field ${\varphi}$, we have
(see, for instance, Appendix A of Ref. \cite{shelton}):
$J_{0R,L}^z = \frac{1}{\sqrt{2\pi}}\; \partial_x \varphi_{R,L}$, $ \;
J_{0R,L}^{+} = \frac{1}{2\pi a_0}\; \re^{\mp \ri \sqrt{8\pi}
\varphi_{R,L}}$.
On the other hand,
combining two Majorana fields, $\xi^1$ and $\xi^2$ to form a Dirac
fermion,
$(\xi^2 + i \xi^1)/\sqrt{2}$, and then bosonizing it, we can express
the SU(2)$_2$ current ${\bf I}$
in terms of a bosonic  field $\Phi$ and the Majorana fermion $\xi^3$:
$I^z _{R,L} = \frac{1}{\sqrt{\pi}}\; \partial_x \Phi_{R,L}$,   
$I_{R,L}^{+} = \frac{\ri}{\sqrt{\pi a_0}}\;\xi_{R,L}^3 \kappa
\re^{\mp \ri \sqrt{4\pi} \Phi_{R,L}}$.
An additional fermionic zero-mode operator $\kappa$ has been introduced
to ensure the correct commutation relations.
Then the Hamiltonian $\bar{\cal H}$
can be written in the following bosonized form:
\bea
\bar{\cal H} &=& v_0 \; [\left(\partial_x \varphi_R \right)^2 +
\left(\partial_x \varphi_L \right)^2 ] +
v_1 \; [ \left(\partial_x \Phi_R \right)^2 +
\left(\partial_x \Phi_L \right)^2 ]\nonumber\\
&-&\ri \frac{v_1}{2} \;
\left( \xi_R^3 \partial_x \xi_R^3  - \xi_L ^3 \partial_x \xi_L ^3 \right)
\nonumber\\
&+& \frac{g_{\parallel}}{\sqrt 2 \pi} \; \left( \partial_x \varphi_L
\partial_x \Phi_R   + \partial_x \varphi_R
\partial_x \Phi_L \right) \nonumber \\
&+& \frac{\ri g_{\perp}}  {2 \left(\pi a_0\right)^{3/2}}
\; \xi_R^3 \kappa \cos \left(\sqrt{4\pi} \Phi_R + \sqrt{8\pi}
\varphi_L\right) .
 \nonumber \\
&+&  \frac{\ri g_{\perp}}  {2 \left(\pi a_0\right)^{3/2}}
\xi_L^3 \kappa \cos \left(\sqrt{4\pi} \Phi_L +
\sqrt{8\pi} \varphi_R
\right).
\label{h1boso}
\eea
We now  perform a canonical transformation:
\bea
\varphi &=& {\rm ch} \alpha \; \bar \Phi_2 + {\rm sh} \alpha \; \bar
\Phi_1, \;
\Phi = {\rm ch} \alpha \; \bar \Phi_1  + {\rm sh} \alpha \; \bar \Phi_2
 \nonumber \\
\vartheta &=& {\rm ch} \alpha \; \bar \Theta_2 - {\rm sh} \alpha \; \bar
\Theta_1, \;
\Theta = {\rm ch} \alpha \; \bar \Theta_1  - {\rm sh} \alpha \; \bar
\Theta_2
\label{cano}
\eea
where $\vartheta$ and $\Theta$ (respectively $\bar \Theta_1$ and $\bar
\Theta_2$)
are the dual fields
associated with $\varphi$
and $\Phi$ (respectively $\bar \Phi_1$ and $\bar \Phi_2$).
The cross terms $\partial_x \varphi \partial_x \Phi$ in Eq. (\ref{h1boso})
can be eliminated by setting
$
{\rm th} 2 \alpha = - g_{\parallel}/\pi \sqrt 2 \left(v_0 +v_1 \right).
$ One immediately observes that
choosing ${\rm th} \alpha = -\frac{1}{\sqrt 2}$, which corresponds to a
special (though nonuniversal) positive value of $g_{\parallel},~(g^*
_{\parallel} =
4 \pi (v_0 + v_1)/3)$,
the arguments of the two cosine terms in (\ref{h1boso})
become those of free fermions, $\cos\left(\sqrt{4\pi} \bar
\Phi_{2L,R}\right)$.
Introducing a pair  of Majorana fields, $\eta$ and $\zeta$, and using
the correspondence: $\psi_{R,L} = (\eta_{R,L} + \ri \zeta_{R,L})/\sqrt{2}
= \;
           (\kappa / \sqrt{2\pi a_0}) \;
\re^{\pm \ri \sqrt{4\pi} \bar \Phi_{2;R,L}}$,
we
finally obtain:
\bea
\bar{\cal H} &=& \frac{u_1}{2} \left[  (\partial_x {\bar \Phi}_1)^2 +
 (\partial_x  {\bar \Theta}_1)^2 \right] - \frac{\ri u_2}{2}
 \left[ \zeta_R\partial_x \zeta_R - \zeta_L\partial_x
\zeta_L\right]\nonumber\\
&-& \frac{\ri v_1}{2}\left[ \xi_R^{3}\partial_x \xi_R^{3}
-\xi_L^{3}\partial_x \xi_L^{3}\right]
- \frac{\ri u_2}{2}
 \left[ \eta_R\partial_x \eta_R - \eta_L\partial_x
\eta_L\right]\nonumber\\
 &+&\ri m \left[ \xi_R^{3}\eta_L - \eta_R\xi_L^{3}\right].
\label{hfin}
\eea
Here $m = g_{\perp}/2 \pi a_0$, and the two
renormalized velocities, $u_1$ and $u_2$,
are expressed in terms of the surface and
middle chain velocities $v_1$ and $v_0$ as  
$
u_1 = (2 v_1 - v_0)/3,\ \  u_2 = (2 v_0 - v_1)/3.
$   
The first two terms
in Eq. (\ref{hfin}) describe
completely decoupled free massless bosonic and Majorana fields,
$\bar \Phi_1$ and $\zeta$, contributing to criticality
with the central charge:
$
C = 1 + 1/2 = 3/2.
$
Therefore, at
the new critical point $\bar{\cal H}$  effectively
represents a gapless $S=1/2$ spin chain and a
critical Ising model.
Coming back to the
model (\ref{hkondo}) and adding the contribution
of the singlet Majorana
fermion $\xi^0$, $i.e.$ one more critical Ising model,
the total central charge becomes $C_T=2$.
The remaining part of the Hamiltonian (\ref{hfin}) has a spectral gap $m$
and  
describes hybridization of the Majorana $\xi^3$ and $\eta$ fields with 
different chiralities.
Since the canonical transformation (\ref{cano}) does not mix ${\cal H}_1$
and
${\cal H}_2$,
the Hamiltonian (\ref{hfin}) still decouples into two commuting, chirally
asymmetric  parts.
This reflects the chiral nature of the fixed point.

{\bf Physical picture of elementary excitations}.
There are two different kinds of elementary excitations 
at the Toulouse point: 
magnetic excitation described   
by the field $\bar{\Phi}_1$, and nonmagnetic, singlet, excitations
associated with the two Majorana fermions $\xi^0$ and $\zeta$.

Notice that, due to the mixing of different degrees of freedom
reflected in the canonical transformation (\ref{cano}),
the ``physical'' spinons, $i.e.$ those defined as
$\sqrt{\pi/2}$-kinks of the field $\bar{\Phi}_1$ describing the effective
$S = 1/2$ spin chain,
should not be misleadingly identified as
the spinons of the middle chain. To get a better understanding of the
structure of
spin excitations at the chiral fixed point, let us express
the currents $J^z _1,~J^z _2$ and $J^z _0$
in terms of the ``physical'' current ${\cal J}^z = (1/\sqrt{2 \pi}) \p_x
\bar{\Phi}_1$. Using the transformation (\ref{cano}) at the Toulouse
point, we
find:
\bea
J^z _{1,2R(L)} &=& {\cal J}^z _{R(L)} - \frac{\ri}{2}
\left(\eta_{L(R)} \zeta_{L(R)} \mp \xi^3 _{R(L)} \xi^0 _{R(L)} \right),
\nonumber\\
J^z _{0R(L)} &=& - {\cal J}^z _{L(R)} + \ri \eta_{R(L)} \zeta_{R(L)}.
\label{curr}
\eea
At energies $|\omega| \gg m$, where all Majorana fields can be considered
as
massless, Eqs. (\ref{curr}) transform back to the
standard definitions of the currents
of the three decoupled chains. In this (ultraviolet) limit, one has a
picture
of three groups of independently propagating spinons. However, in the IR
limit ($|\omega| \ll m$), all Majorana bilinears in
(\ref{curr}) are characterized by short-ranged correlations (since the
Majorana fermions $\eta$ and $\xi^3$ are massive),
implying that strongly fluctuating
parts of the currents of individual chains are no longer independent; all
of
them contribute to the formation of a single, physical, current ${\cal
J}^z$.
In fact,
the physical spinon represents a chirally asymmetric,
strongly correlated
state of {\it three} spinons.
Consider, for instance, a right-moving
$\sqrt{\pi/2}$-kink of the field $\bar{\Phi}_1$, representing a physical
spinon with the spin projection $S^z = 1/2$. According to the exact
relation
\be
{\cal J}^z _{R(L)} = J^z _{1R(L)} + J^z _{2R(L)} + J^z _{0L(R)}
\label{rel}
\ee
following from (\ref{curr}), such
an excitation is a combination of two right-moving spinons of the surface
chains, each carrying the spin $S^z = 1/2$, and a left-moving antispinon
of the
middle chain, with $S^z = - 1/2$.
The rigidity of such a state is ensured by a finite mass gap in the
$(\eta-\xi^3)$
sector of the model.   
This peculiar structure of the elementary
spin excitations at the chiral fixed point is also reflected by the
expression
for the velocity $u_1$.

Apart from the nontrivial nature of the spinon, the chiral
fixed point manifests
itself in the existence of two new massless singlet excitations.
The Majorana
fermion $\xi^0$ describes collective
excitations of singlet pairs formed on the two surface chains. The
nature of the
Majorana fermion $\zeta$ is less transparent: it is a highly nonlocal
object when expressed
in terms of the original spin operators. The best 
way to understand the
role
of the singlet excitations is to combine the $\xi^0$ and $\zeta$ fields
into a single Dirac
fermion and then bosonize it.
The corresponding
massless
bosonic field ${\tilde \Phi}_c$ resembles
the scalar field describing the charge
degrees of freedom in the Hubbard model away from half filling
in the limit  $U=\infty$.  We shall hence 
refer to the gapless Majorana fields as
``pseudocharge'' excitations which account for
the central charge $C=1$.
The spin-pseudocharge separation is already
manifest at the Toulouse point implying that
the leading asymptotics of  correlations functions
will factorize into the spin and pseudocharge contributions.

{\bf Correlation functions}.
We shall now use the exact solution of the model,
found at the Toulouse point,
to calculate physical quantities of interest. From Eqs. (\ref{curr}) it
follows
that, as for a single $S = 1/2$
Heisenberg spin chain, an external magnetic field
$H$ couples only to the massless field $\bar  \Phi_1$.
%${\cal H}_{\em mag} = - H /\sqrt{2 \pi})\p_x \bar  \Phi_1$.
The uniform susceptibility in units of $g\mu_B$
is easily found to be $\chi^{-1} = 2 \pi u_1$.
At low temperatures ($T \ll m$) only the gapless modes ($\xi^0,~ \zeta$
and
$\bar{\Phi}_1$) contribute to
the specific heat.
Using the general formula $C_V = \pi C T/3v $,
we find: ${\displaystyle
 C_V = \frac{\pi T}{3} \left( \frac{1}{2} \cdot \frac{1}{v_1}
 + \frac{1}{2}\cdot  \frac{1}{u_2}
+ 1 \cdot  \frac{1}{u_1}\right)}$.

We have also computed all spin-spin correlation functions at the Toulouse
point (details will be published elsewhere\cite{azaria}).
The leading asymptotics of the uniform part of the correlation functions
coincide with that of the effective $S = 1/2$ spin chain,
in agreement with the above
discussion.
The nature of the chiral
fixed point manifests itself in {\it staggered} correlations
 between the spins of the
surface chains:
\bea
&&\langle {\bf  n}_a\left(x,\tau\right) {\bf n}_b\left(0,0\right)\rangle
\nonumber \\
&\sim&
  \delta_{ab} \frac{1}{ (x^2 +
u_1^2\tau^2)^{1/2} } \frac{1}{(x^2 + v_1^2
\tau^2)^{1/8}}\frac{1}{(x^2 + u_2^2
\tau^2)^{1/8}} \label{n-n}
\eea
where $a, b =(1,2)$. Notice the factorized contribution
of the ``pseudocharge''
coming with the exponent $1/8$.
As seen from Eq. (\ref{n-n}), there is an intrinsic velocity
anisotropy in the pseudocharge sector 
which might be important when
considering
the dynamical properties. We emphasize that the exponents in the
correlation
functions are {\it universal} and 
characterize  a new universality class
in
spin ladders. This is the main result of our work. Another quantity of
experimental
interest is the NMR relaxation rate $1/ T_1$. It is not difficult
to show that at low temperature, $1/ T_1 \sim \; \sqrt T$,
in contrast with the Heisenberg chain where $1/ T_1 \sim \;  \rm const$.

{\bf Stability of the chiral fixed point}.
With all these results at hand,
now we turn to the stability
of the chiral fixed point. There are two important
questions we shall adress.
The first is related to the stability of the Toulouse point with
$g_1 = 0$ kept fixed,
and the second is to examine
the behavior of the system when one moves
away from the point $J_{\perp} = 2 J_{\times}$.

We begin by  stressing that our solution is stable
provided both $u_1$ and $u_2$ are positive, the stability condition
thus being
$
1/2 \le v_0/v_1 \le 2.
$
So the chiral fixed point is stable in
a relatively broad range of velocities.
When $g_{\parallel}$ deviates from its Toulouse-point value, the
Hamiltonian
(\ref{hfin}) picks up an extra term
$\delta g_{\parallel}  (\partial_x {\bar \Phi}_{1L} \zeta_R \eta_R$
$+ \partial_x {\bar \Phi}_{1R} \zeta_L \eta_L )$.
Since the field
$\eta$ is massive, the expansion in $\delta g_{\parallel}$
does not introduce new infrared singularities, implying  
that
the long-distance behavior of the correlation functions  
will not be modified except for a velocity and mass renormalization.
Therefore, 
the solution at the Toulouse point captures all universal properties of
the
chiral fixed point and includes the case  $g_{\parallel} = g_{\perp}$
where
the Hamiltonian has the full SU(2) symmetry.

Now we consider small
deviations from the  point $g_1 = 0$. At small enough $|g_1|$
it is possible to investigate the effect
of the backscattering term
${\bf n}_0 \cdot \left({\bf n}_1 + {\bf n}_2\right)$
as a weak perturbation to the chiral fixed point. This perturbation can be
shown to be proportional
to $\cos(\sqrt{\pi} {\tilde \Phi}_c)$, which is a relevant
operator
with scaling dimension $1/4$.
We thus conclude that
the backscattering operator opens a gap, $\Delta_c$, in the
pseudocharge sector but has no
effect on the magnetic (spinons) excitations.
Standard scaling arguments give an
estimate: $\Delta_c \sim g_1^{4/7}$.
The chiral fixed point is thus unstable in the far IR limit,
and the system will flow to
the $C=1$ fixed point
of the standard  three-leg spin ladder.
Of course, the very applicability of the perturbative approach
to the chiral fixed point requires
that $\Delta_c \ll m$,
the condition which can always be satisfied for sufficiently small $g_1$.
Under this condition, there exists an
intermediate but still low-energy region
$\Delta_c \ll E \ll m$ where
the $C=2$ behavior caused by frustration is dominant.
The physics in this region is
universal and cannot be
understood without having recourse to the new chiral
fixed point. 
At lower energies, $ E \ll \Delta_c$, the system will eventually
cross over to the conventional
critical $C=1$ behavior.

We think that this new fixed point
discussed on the present paper 
might be responsible for new physics in many
frustrated ladders such as the
three-chain zigzag ladder. 
This model and the doped case  are currently under study.
We hope that the chiral-fluid critical state
with all its physical properties 
will be
observed in further experiments on spin-ladder systems.

The authors would like to thank
D. Allen, N. Andrei, F. Essler, R. Flume, A. Gogolin,
C. Lhuillier, A. M. Tsvelik
and Lu Yu for very helpful conversations.
A.N. acknowledges the support from Deutsche Forschungsgemeinschaft.
He would also
like to thank Vladimir Rittenberg for his kind hospitality and stimulating
discussions.

\sloppy
\par

%\newpage


\begin{thebibliography}{99}
\bibitem{haldane} F. D. M. Haldane, Phys. Lett. {85A}, 375 (1981).
\bibitem{dagotto} See, for a review, E. Dagotto and T. M. Rice,
Science {\bf 271}, 618 (1996), and references therein.
\bibitem{khve} D. V. Khveshchenko, Phys. Rev. B {\bf 50}, 380 (1994).
\bibitem{coldea} R. Coldea {\it et al}, Phys. Rev. Lett. {\bf 79}, 151
(1997).
\bibitem{chab}
G. Chaboussant {\it et al}, Phys. Rev. B {\bf 55}, 3046 (1997).
\bibitem{white} S. White and I. Affleck, Phys. Rev. B {\bf 54},
9862 (1996).
\bibitem{allen} D. Allen and D. S{\'e}n{\'e}chal,
Phys. Rev. B {\bf 55}, 299 (1997).
\bibitem{shura}  A. A. Nersesyan, A. O. Gogolin, and
F. H. L. Essler, to appear in Phys. Rev. Lett., cond-mat/9804005.
\bibitem{oitmaa} Z. Weihong, V. Kotov, and J. Oitmaa, 
Phys. Rev. B {\bf 57}, 11439 (1998); X. Wang, cond-mat/9803290.
\bibitem{arri} E. Arrigoni, Phys. Lett. A {\bf 215}, 91 (1996);
K. Totsuka and M. Suzuki, J. Phys. A: Math. Gen. {\bf 29}, 3559 (1996);
D. Schmeltzer and P. Sun, J. Phys.: Condens. Matter {\bf 10}, 4435 (1998).
\bibitem{andrei} N. Andrei, M. Douglas,
and A. Jerez, cond-mat/9803134.
\bibitem{affleck} I. Affleck, Nucl. Phys. B{\bf 265}, 409 (1986).
\bibitem{azaria} P. Azaria, P. Lecheminant, and A. A. Nersesyan,
in preparation.
\bibitem{shelton} D. G. Shelton, A. A. Nersesyan, and A. M. Tsvelik,
Phys. Rev. B {\bf 53}, 8521 (1996);
A. A. Nersesyan and A. M. Tsvelik, Phys.
Rev. Lett. {\bf 78}, 3939 (1997).
\bibitem{emery} V. J. Emery and S. A. Kivelson, Phys. Rev. B {\bf 46},
10 812 (1992).
\bibitem{polyakov} A. M. Polyakov and P. B. Wiegmann, Phys. Lett. B {\bf
141}, 223 (1984).


\end{thebibliography}
\end{document}